# Laser-assisted generation of elongated Au nanoparticles and subsequent dynamics of their morphology under pulsed irradiation in water and calcium chloride solutions

M.I. Zhilnikova[1,2,*], E.V. Barmina[2], G.A. Shafeev[2,3], A.V. Simakin[2], S.M. Pridvorova[4], O.V. Uvarov[2]

1 The Federal State Educational Institution of Higher Professional Education, "Moscow Institute of Physics and Technology (State University)", 9, Institutsky lane, 141700, Dolgoprudny, Moscow region, Russian Federation

2 Wave Research Center of A.M. Prokhorov General Physics Institute of the Russian Academy of Sciences, 38, Vavilov street, 119991 Moscow, Russian Federation

3 National Research Nuclear University MEPhI (Moscow Engineering Physics Institute), 31, Kashirskoye highway, 115409, Moscow, Russian Federation

4 A.N. Bach Institute of Biochemistry, Federal Research Centre «Fundamentals of Biotechnology» of the Russian Academy of Sciences, 33, Leninsky prospect, 119071, Moscow, Russian Federation

*MZhilnikova@gmail.com

## Abstract

One-step laser generation of Au elongated nanoparticles (NPs) and their successive fragmentation and agglomeration are experimentally studied for the first time. In the present work, laser-assisted formation of Au elongated nanoparticles by ablation of a solid Au (99.99%) target in water was done using a ytterbium-doped fiber laser sources with pulse duration of 200 ns and pulse energy of 1mJ. Extinction spectrum correlating with TEM shows the appearance of absorption signal in red region and near IR-spectrum that corresponds to longitudinal plasmon resonance of electrons in elongated Au NPs. In addition, generated elongated Au nanoparticles were exposed to pulsed laser beam with various pulse energy and laser exposure time**.** It was found that at early stages of irradiation NPs agglomerate as the NPs chains with size of order of 1 µm long. Further laser exposure results in fragmentation of these chains. Possible processes of laser-assisted formation of elongated Au NPs in aqueous solutions of calcium chloride and their subsequent interaction with pulsed laser irradiation are discussed.

## Introduction

Laser ablation of solids in liquids is a technology for generation of nano-objects [1−5]. Laser-assisted formation of nanoparticles is the result of laser beam impact on a target and subsequent removal of material from the target surface, promoted the liquid surrounded by vapor. At sufficiently high laser fluence, the target surface layer (solid at room temperature) undergoes melting. The medium around the target, which is liquid under normal conditions, passes to the overheated gaseous state. The vapor at high pressure interacts with the molten layer on the target surface dispersing it into surrounding liquid as nanoparticles. Laser ablation in liquid is used to synthesize large variety of nanoparticles. A lot of attention is paid now to this technique because of its simplicity and possibility of chemically pure nanoparticles formation [6−8].

To date, laser-assisted formation of Au nanoparticles has been investigated in numerous studies, because this metal is chemically inert [9, 10] and chemical interaction of ejected NPs of Au with overheated liquid vapors is excluded. A number of studies [11−13] were devoted to the processes of generation and optical properties of Au NPs. Like many other NPs generated by laser ablation in liquids, individual Au NPs may interact with the laser beam inside the liquid, which leads to change of their morphology and size distribution function.

Until now, the effect of laser radiation on gold nanoparticles has been studied for only spherical particles, which may behave differently: undergo fragmentation [14−17,18] or agglomeration [10, 19, 20]. Previously, it was shown both experimentally and theoretically that the main fragmentation mechanism for nanoparticles less than 100 nm in size is the detachment of small fragments of order of 10 nm in size [21]. If the initial colloid contains larger nanoparticles (∼10 to 1000 nm in size), the separation of micro-particles into halves (fragments of approximately equal sizes) becomes important [22]. Agglomeration of nanoparticles was observed under laser irradiation of colloids with a high nanoparticles concentration [23] or when the particles had some additional charge. This charge was provided by an external electric field and/or the presence of β-active impurities, such as Tritium [24]. Thus, depending on the experimental parameters, laser irradiation of colloidal systems may lead to either fragmentation or agglomeration of initially spherical NPs. Sphere is the surface that provides the minimum of surface energy of the melt. That is why laser-generated NPs are usually spherical.

Experiments on laser-assisted formation of elongated gold nanoparticles under irradiation of a colloidal solution of spherical metal nanoparticles by IR laser radiation with a high (35 to 80 mJ) pulse energy were performed in [10]. However, the formation of elongated gold nanoparticles specifically as a result of laser ablation of a gold target has not been observed until now. Usually, the water used in ablation experiments is either pure or may contain intentionally added ionic additives. Studies have been conducted on the formation of gold nanoparticles by laser ablation in solutions of sodium chloride and potassium chloride [27], i.e. single valence cations. New feature is observed in our case where additionally added is calcium chloride, i.e. two-valence cation**.** Their presence alters the dynamics of laser interaction with Au NPs in a qualitative way.

The purpose of this study is the single-step generation of elongated Au nanoparticles by laser ablation of a solid gold target in aqueous solutions of Calcium chloride salts and their subsequent investigation of size distribution dynamics under Au colloid interaction with pulsed laser radiation.

## Experimental

Gold nanoparticles were obtained by laser ablation of a solid target in liquid. To this end, radiation of an ytterbium-doped fiber laser (pulse width of 200 ns, the repetition rate of 20 kHz, pulse energy of 1 mJ) with a wavelength of 1060 to 1070 nm was focused by an objective (F-Theta = 21 cm) onto the surface of a gold (purity of 99.99%) plate immersed in water prepared by reverse osmosis equipped with additional filter (12 ml in volume). The water contained some amount of impurities: sodium, potassium, and calcium salts. The overall salts content corresponded to that of drinking water. The diameter of the laser-beam focal spot on the target was about 50 μm. The sample surface was scanned by a focused beam using a galvo mirror system (2 pulses per spot). The thickness of the liquid layer above the target was 2 to 3 mm. Laser exposure time was 1 min.

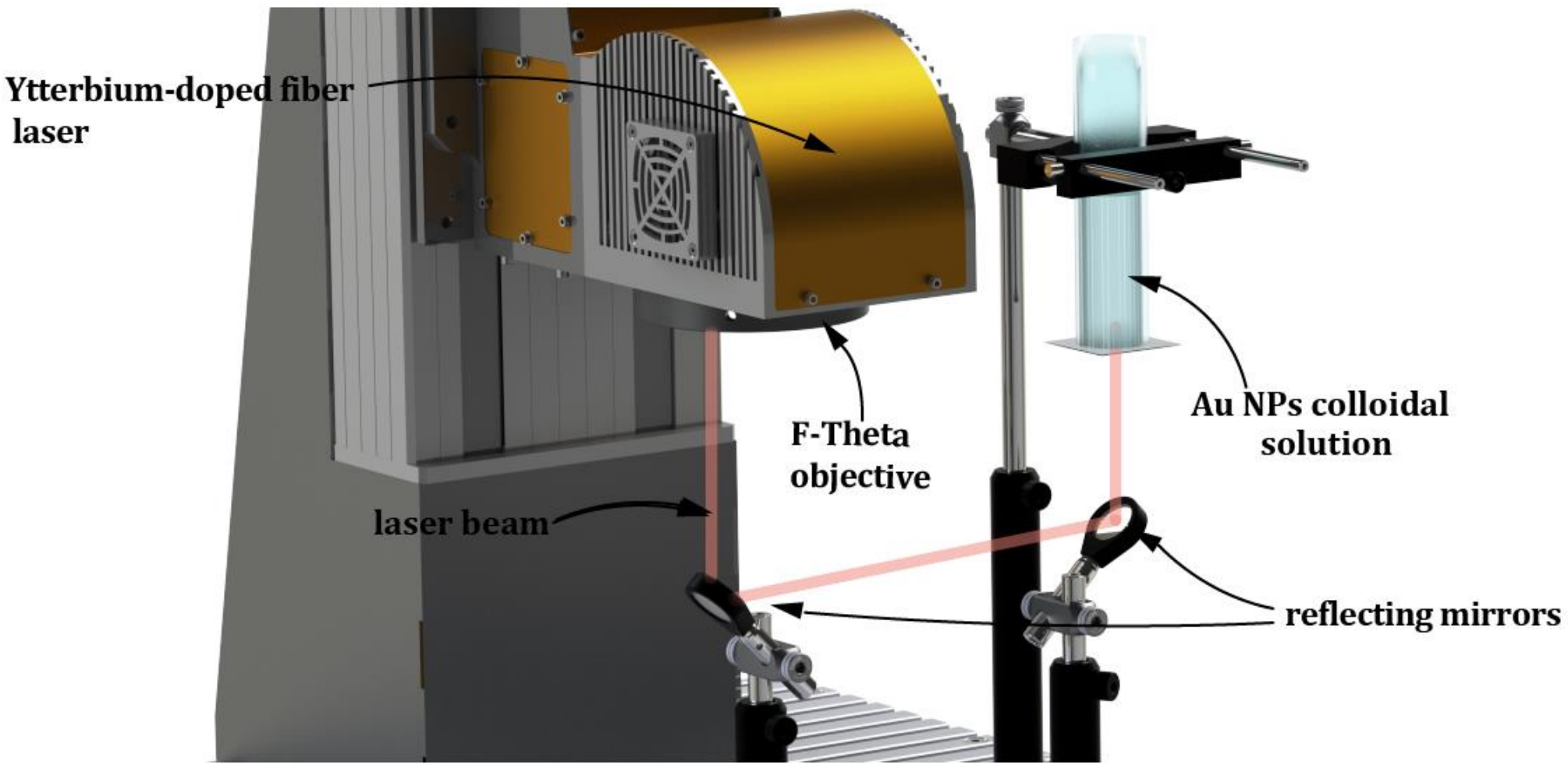

**Fig.1.** Schematic of the experiment on laser irradiation on colloidal solutions of elongated gold nanoparticles.

Second part of experiments has been devoted to investigation of the laser interaction with colloidal solutions of elongated nanoparticles. In this case the beam of ytterbium-doped fiber laser was passed through two reflecting mirrors making an angle of $45^{\circ}$ and focused in the upward direction through a flat window onto the bottom of a glass cell filled with previously prepared colloidal solution (Fig. 1). The laser beam waist in the colloid-containing cell was spaced by a distance of 2 to 3 mm from the input window surface to avoid the window damage. The experimental setup favored the convection of colloidal solution and, as a consequence, its active mixing. Laser beam moved along a circle trajectory with linear velocity of 500 mm $s^{-1}$. The irradiation of a colloidal solution was accompanied by plasma formation due to the optical liquid breakdown. Two parameters were varied in all experiments: irradiation time (from 1 to 60 min) and pulse energy (1 and 0.5 mJ). Thus, the estimated fluence in the solution was either 40 or 20 J $cm^{-2}$. The volume of colloidal solution was 3 ml at each laser exposure. The extinction spectra of the thus prepared colloids were measured in the range of 200 to 900 nm using an Ocean Optics UV-vis fiber spectrometer and from 400 nm to 1800 nm by Shimadzu UV-3600 spectrometer. The nanoparticles morphology was analyzed with a Carl Zeiss 200FE transmission electron microscope (TEM).

## Results

Gold nanoparticles were formed using nanosecond pulsed laser radiation via ablation of a solid gold target in water. The extinction spectrum of the initial nanoparticles [28] sample has a peak at wavelengths of 520 nm, corresponding to the transverse plasmon resonance of spherical gold nanoparticles. A comparative analysis of this spectrum with the theoretical spectrum for spherical nanoparticles [25] revealed absorption in the red spectral region [17]. Due to this, the colloidal solution of elongated gold nanoparticles appeared blue in appearance rather than red, as in the case of spherical nanoparticles. The TEM image of the initial sample (Fig. 2) exhibits individual chains of nanoparticles (they are encircled in the image).

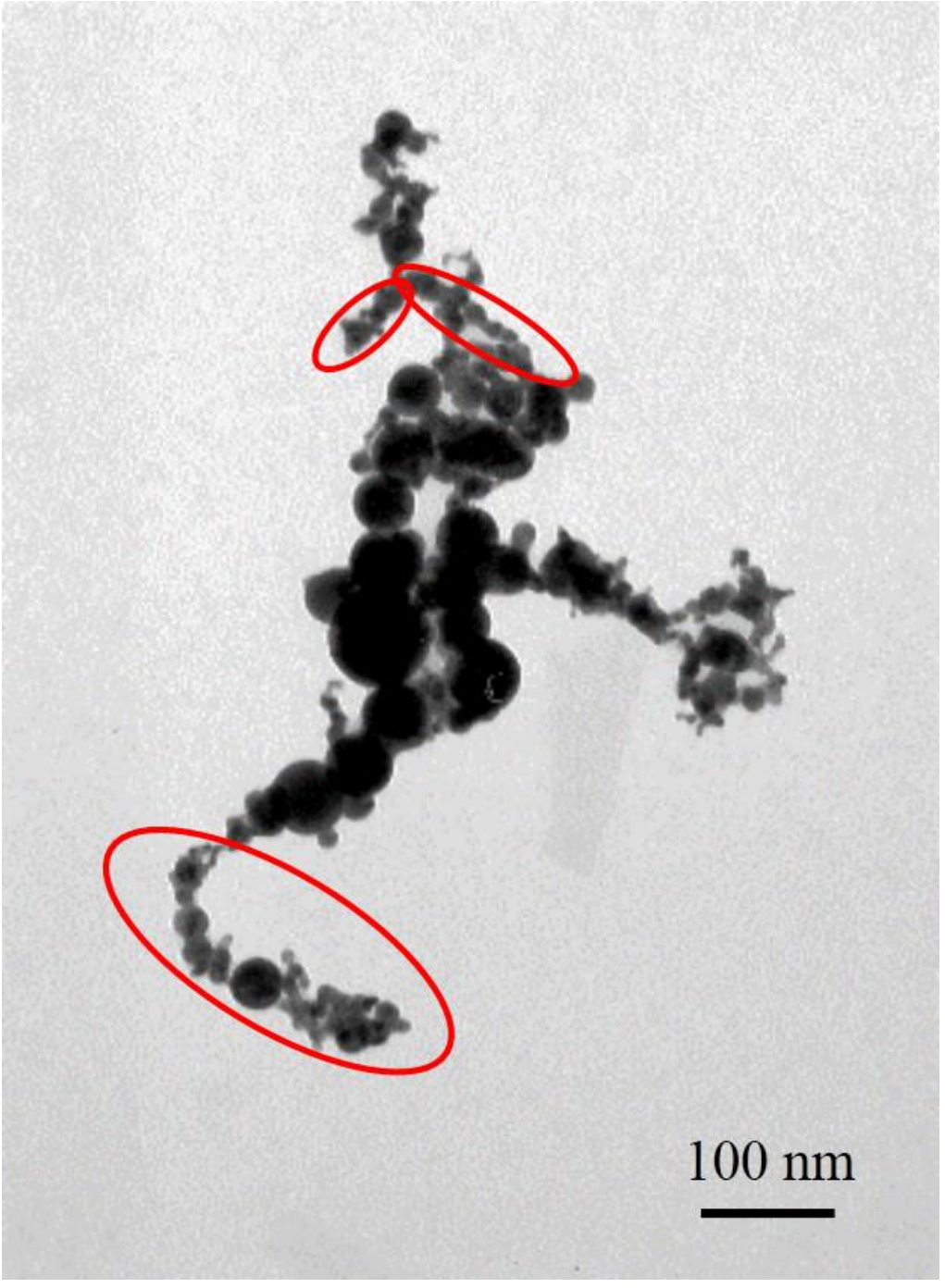

**Fig.2**. TEM image of gold NPs formed by ablation of a gold target in water using an ytterbium-doped fiber laser. Chains of nanoparticles are encircled in red ovals.

It is known [17] that the existence of these chains leads to absorption in the red spectral region: their elongated shape causes longitudinal plasmon resonance, i.e., oscillations of free electrons in nanoparticles along their longer axis.

To investigate the morphology dynamics of the elongated nanoparticles during their laser irradiation exposure both time and pulse energy were varied.

In the first series of experiments, the pulse energy was constant (1 mJ), whereas the irradiation time varied from 1 to 60 min. Fig. 3 shows the dependence of optical density of the colloidal solution in the red region on the laser exposure time. Absorption in the red region of spectrum corresponds to presence of elongated Au NPs. Exposure time 0 corresponds to the initial colloid generated by laser ablation of the bulk target.

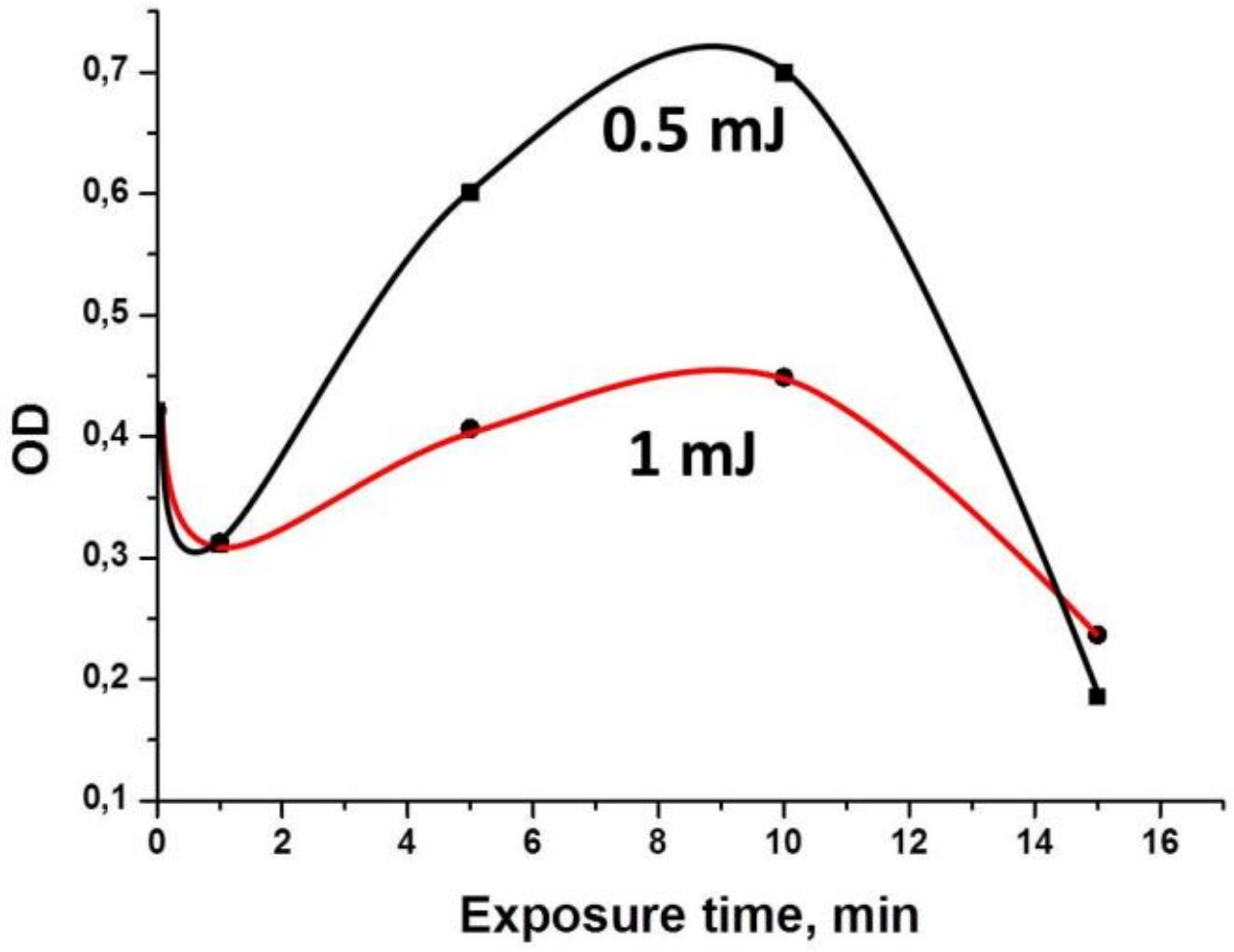

**Fig.3.** Dependence of the optical density of solutions on the laser irradiation time at a wavelength of 700 nm with pulse energy of 0.5 and 1 mJ.

One can see that the optical density of colloids in the red region first increases with laser exposure time and then passes through maximal value and decreases. The decrease of OD indicates the decrease of concentration of elongated Au NPs. Note that the absorption in red region is higher at lower laser pulse energy 0.5 mJ.

Fig.4 demonstrates typical morphology of elongated nanoparticles after their laser exposure during 5 min.

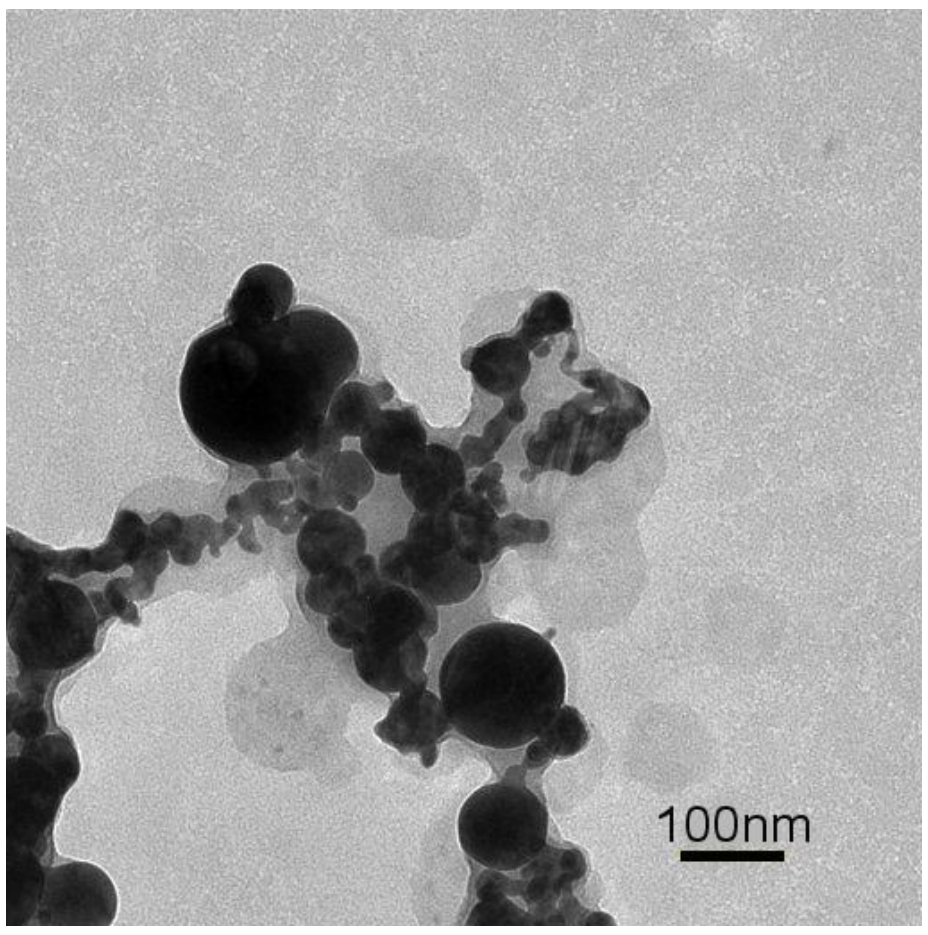

a

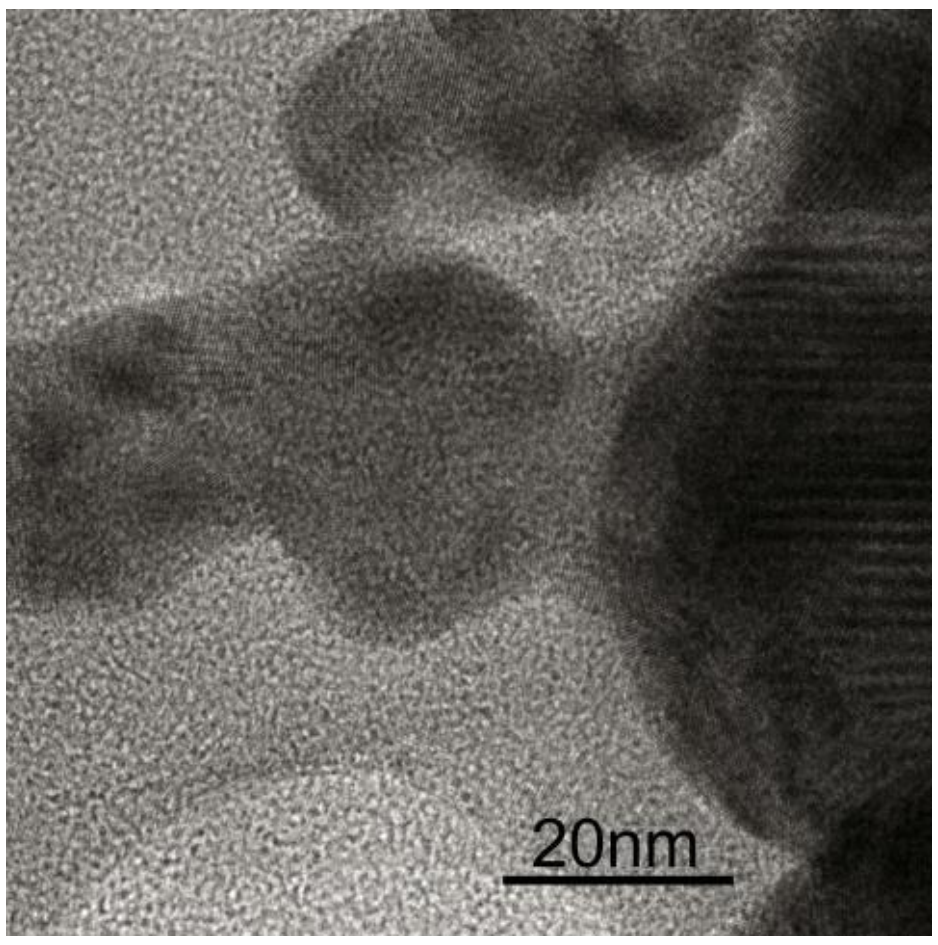

b

**Fig.4.** (a) TEM image of gold nanoparticles after laser irradiation of its colloidal solution at an exposure time of 5 min and pulse energy of 1 mJ. Scale bar denotes 100 nm. (b) HRTEM image

of elongated Au NPs. One can see crystallographic planes of Au and the part of the low-contrast shell around Au NPs. Scale bar denotes 20 nm.

As can be noticed in Fig. 4(a), 5-min laser irradiation (pulse energy 1 mJ) of a colloidal solution leads to the formation of large nanoparticles (about 100 nm in size), linked by nanoparticles chains, whose length reaches 150 nm. Isolated diffuse particles outside Au NPs are also visible. Fig. 4(b) shows the High Resolution TEM view of Au NPs. One can see atomic planes of Au, while the diffuse halo around them contains no crystalline structure.

In the second series of experiments, the average pulse energy was 0.5 mJ. The evolution of the absorption spectrum of the colloidal solution in the VIS-IR range of Au NPs depending on the laser exposure time is shown in Fig.5.

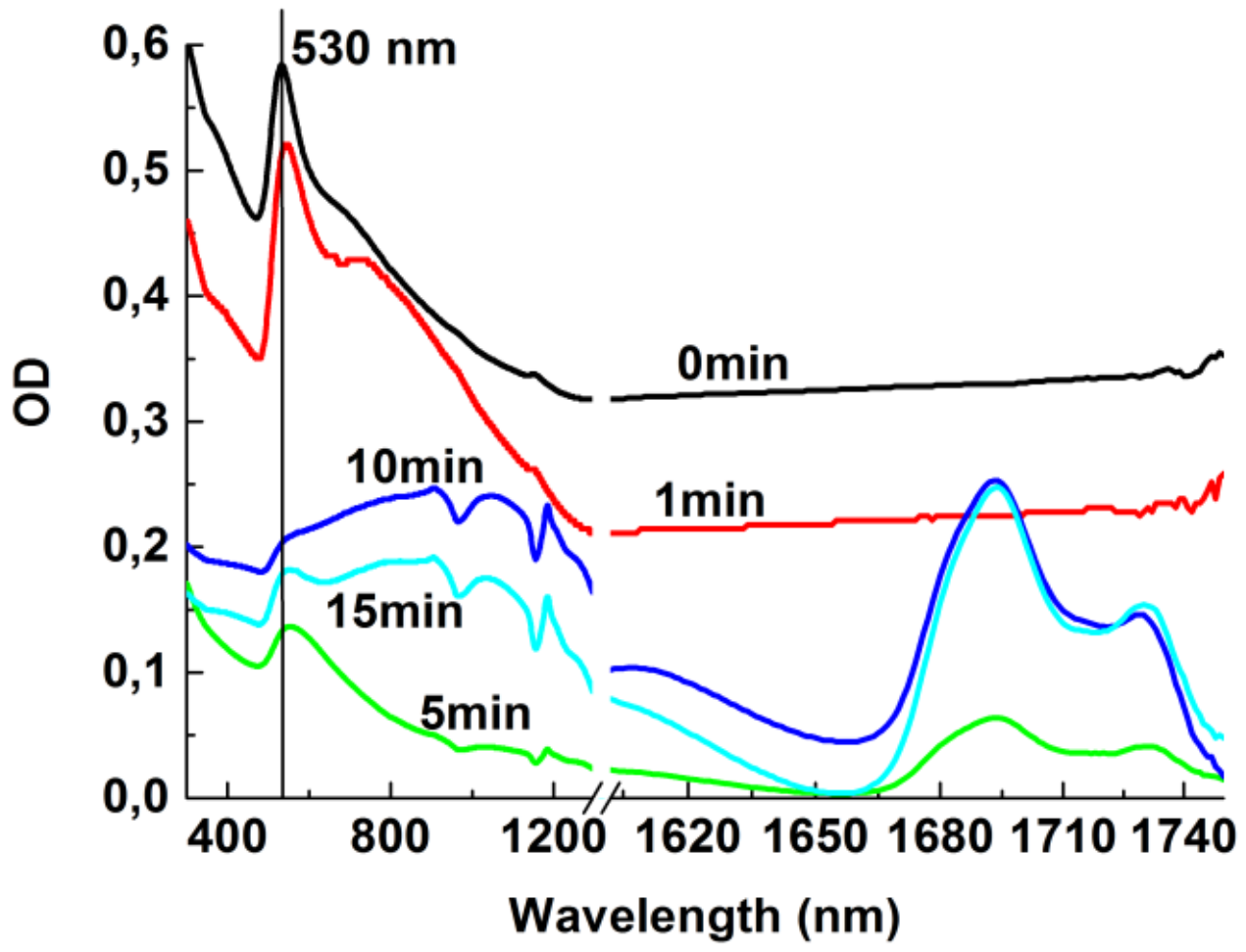

**Fig.5.** Dependence of Au NPs extinction spectrum in the UV-IR range on laser exposure time. The pulse energy is of 0.5 mJ.

In this case, an important fact is that there are significant absorption peaks in the IR range at 10 and 15 minutes of laser exposure time, which indicates the presence of elongated nanoparticles (nanorods).

Water used in previous set of experiments presented above contained some salts. Energy Dispersive X-ray analysis (EDX) of residual of evaporation of many drops of this water showed the presence of Na, K, Ca, and Cl in water. In a series of previous experiments we found that the addition of NaCl to pure water in concentration up to 100 mg/ml did not result in the formation of elongated nanoparticles. By assumption the presence of calcium ions $Ca^{2+}$ in the water contributes to the formation of elongated Au NPs. Concentration of $CaCl_2$ in the water used was 0.88 mg/l, as measured by flame atomic absorption spectroscopy.

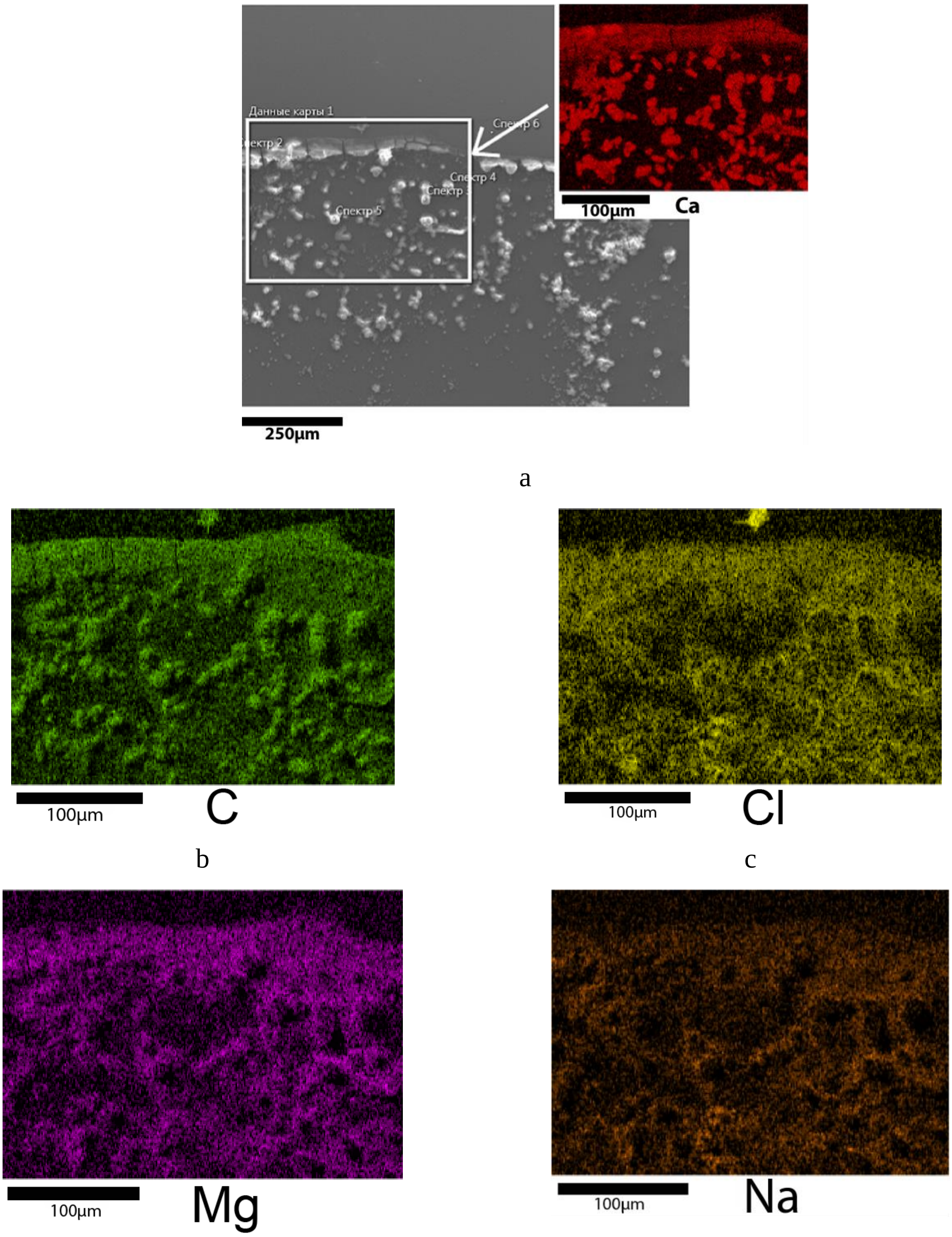

d e

**Fig.6.** TEM image of evaporated water droplets on a silicon substrate. The inset shows a significant presence of calcium (a). A small amounts of carbon (b), chlorine (c), magnesium (d) and sodium (e) was also found.

Following set of experiments on irradiation of the gold target by laser ablation in the liquid were carried out in water Milli-Q with various concentrations of calcium chloride ($CaCl_2$).

The laser exposure time is 1 min with various concentrations of calcium chloride. The extinction spectra of Au NPs in water obtained by laser irradiation on gold target for various concentrations of $CaCl_2$ are presented in Fig.7. A pronounced shift in the absorption maximum is observed with increasing concentration.

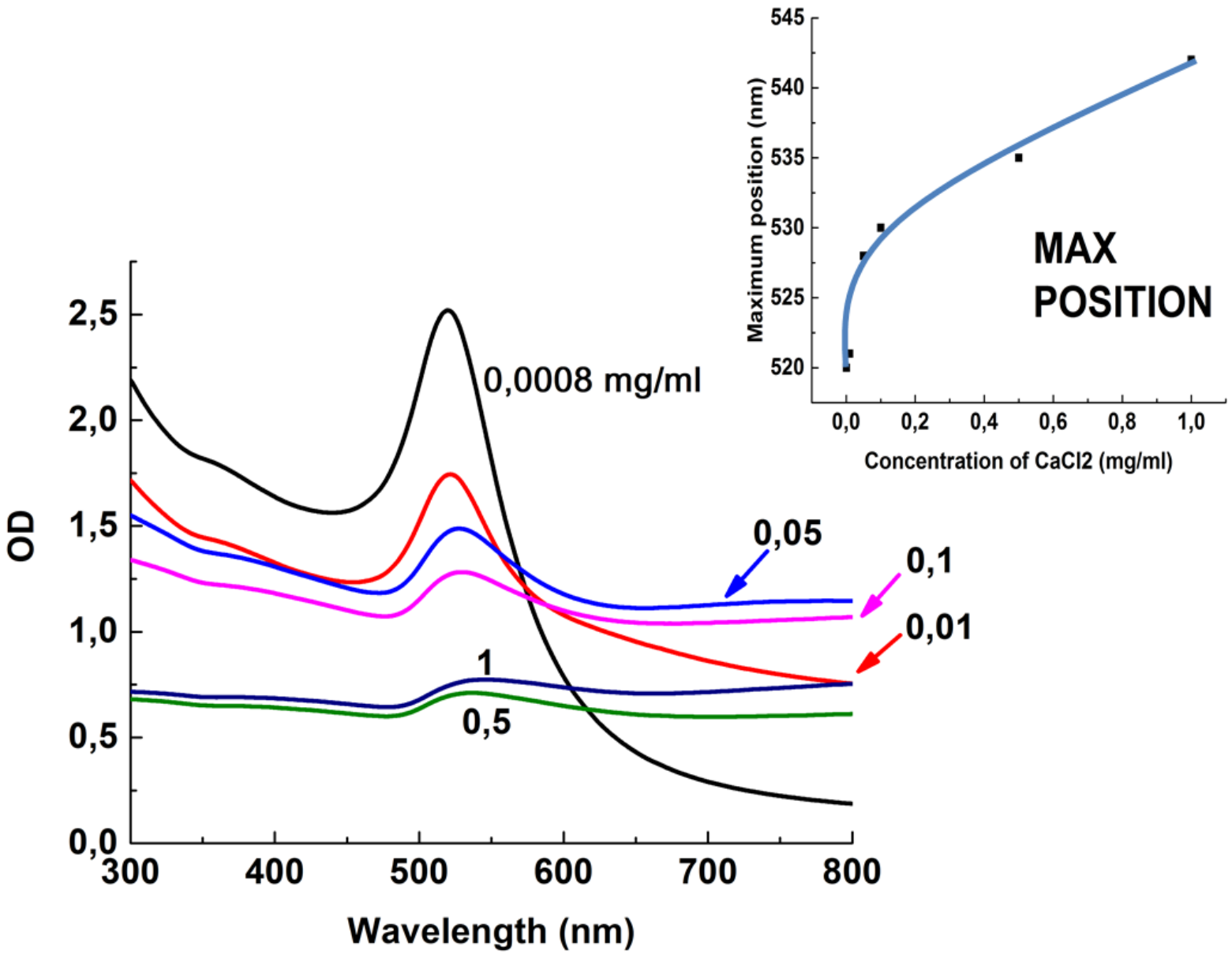

**Fig.7**. Dependence of Au NPs extinction spectra on the concentration of $CaCl_2$ in pure water (Milli-Q). The insert contains the position of the peaks of extinction spectra depending on the concentration. The pulse energy of 1 mJ, the exposure time is 1 min.

The graphs show that at a concentration of 0.1 mg/ml the peak is at 520 nm, and at 1 mg/ml already at 540 nm. According to the assumption, this shift is caused by the increase in the refractive index around gold nanoparticles due to the presence of $Ca(OH)_2$ shell around them [25].

For the analysis of the obtained optical spectra, the extinction spectrum of water was measured with the addition of different concentrations of $CaCl_2$. It was found that calcium chloride does not make a significant contribution to the spectrum for gold nanoparticles (the maximum optical density reaches only 0.03), and therefore the peak at 250 nm corresponds to the absorption of nanoparticles. In addition, it should be noted that the optical density of an aqueous solution of calcium chloride is negligible in the visible region of the spectrum.

At higher concentrations of TEM-image of $CaCl_2$ the Au NPs are also elongated (Fig.8).

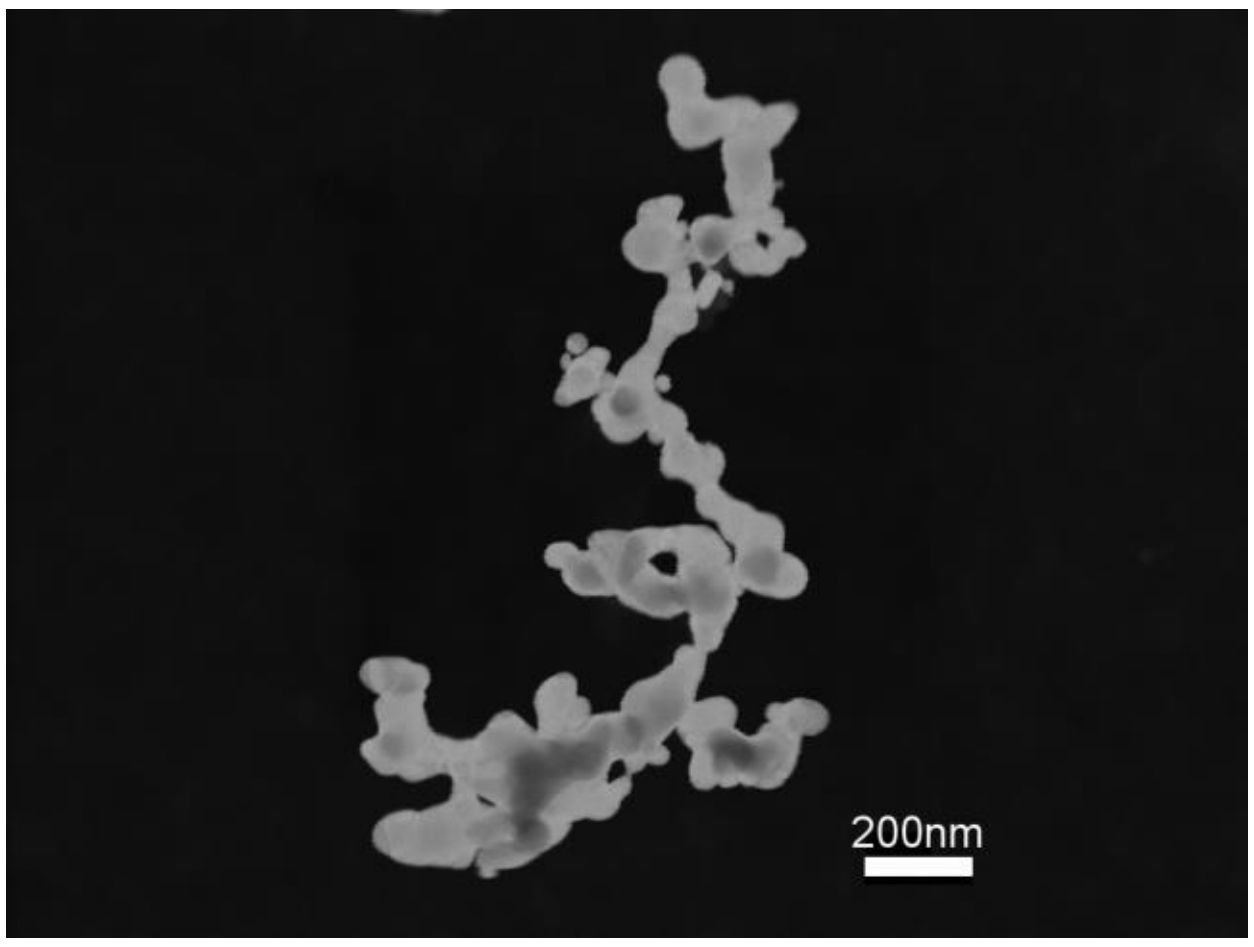

**Fig.8.** TEM image Au NPs in scattered electrons after laser irradiation of colloidal solution Au NPs with the $CaCl_2$ content of 10 mg/ml at an exposure time of 1 min. The pulse energy is of 1 mJ. Scale bar denotes 200 nm.

One can observe the appearance of a chain of elongated Au NPs with a longitudinal size of more than 400 nm.

It is pertinent to note that aqueous solution of $CaCl_2$ may promote the formation of elongated Au NPs even in absence of laser exposure. By addition of this solution at concentrations of 10 mg/ml and 50 mg/ml to Au NPs generated by laser ablation in pure water (milliQ) the shift of the transverse plasmon resonance of 4 nm to the red is observed. More pronounced effect is simultaneous appearance of absorption wing in red region typical of elongated Au NPs. The intensity of this wing however is much weaker than after laser exposure.

## Discussion

As follows from the above results, laser irradiation of a colloidal solution of elongated nanoparticles causes their agglomeration, which is accompanied by increasing absorption in the red and IR spectral regions. The initial elongated nanoparticles have plasmon resonances spaced from the laser wavelength (1060 to 1070 nm). However, during elongation, longer nanoparticles may come to resonance with laser radiation. Apparently, this instant corresponds to the transition from agglomeration to fragmentation of elongated nanoparticles. An extrapolation of data [26] shows that elongated nanoparticles (with an aspect ratio of five to six) absorb at the laser wavelength of 1060-1070 nm.

The presence of a wide wing in the spectrum of gold nanoparticles formed by laser ablation is due to the superposition of longitudinal resonance peaks for NPs with different aspect ratios.

As can be seen in Fig.4 (a) the gold NPs are surrounded by a low-contrast halo, which constitute a dielectric substance with a low atomic mass. It is impurity of calcium salt, which is confirmed by the chemical analysis of the water from experiments. The shift of the main maximum occurs due to the formation of a shell on the nanoparticle itself during the laser action, since it is sensitive to the refractive index of surrounding medium [25].

Aqua-ions $Ca^{2+}$ may acquire electrons from breakdown plasma and be reduced to metallic Ca. The interaction of Ca atoms with water at elevated temperatures is accompanied by the formation of insoluble in water $Ca(OH)_2$. We suggest that it is calcium hydroxide or hydroxide that forms the shell around Au NPs, as mentioned above. Indeed, the absorption spectra of the Au NPs show a positive effect of the presence of calcium ions in water on the formation of elongated particles. TEM-images also confirm this fact.

The transfer of agglomeration to fragmentation of Au NPs observed with the increase of laser exposure time (see Fig. 3) can be assigned to the limited amount of Ca ions in the solution. Indeed, at short exposure times newly formed product $Ca(OH)_2$ dominates over fragmentation of Au NPs leading thus to further elongation of Au chains. However, when the content of Ca ions decreases, then the fragmentation of Au NPs decorated with $Ca(OH)_2$ starts to proceed resulting in the formation of smaller Au NPs along with the decrease of optical density of the colloid in red region of spectrum (Fig. 3). The elongation of Au NPs under addition of aqueous solution of $CaCl_2$ to initially spherical Au NPs can be explained by excessful negative charge of as-generated Au NPs. This charge is also capable of reducing $Ca^{2+}$ ions to lower oxidation state.

Reduction of $Ca^{2+}$ ions to $Ca^0$ and then through interaction with $H_2O$ to $Ca(OH)_2$ leads to formation of the shell around elongated Au NPs but apparently not inside the Au NPs. This stems from the fact that longitudinal plasmon resonance of elongated Au NPs is observed for high aspect ratios. Therefore, the metallic medium along elongated Au NPs is continuous without inclusion of any dielectric species. Weak crystallinity of $Ca(OH)_2$ visible in HRTEM images (see Fig. 4 (b)) suggests that the shell is made of individual molecules of this compound.

Elongation of hematite nanoparticles in presence of $Ca^{2+}$ cations and alginate polymer has been reported in the paper by Kai Loon Chen *et al*, Langmuir 2007, 23, 5920-5928. The chemical composition of the solution in our case is different. The closest to our work is the work about formation of chains of chemically synthesized spherical Au NPs [29]. The authors observed the formation of chains by adding aqueous solution of $CaCl_2$ to the colloid of spherical Au NPs cupped with citrate anions. The process of Au NPs chains formation is governed by the interplay between electrostatic repulsion of Au NPs and their agglomeration due to van der Waals attraction. At certain surface charge of NPs (zeta-potential) there is repulsion from sides of the chain and attraction on its end. This leads to growth of the chain made of spherical NPs and corresponding formation of the wing in the red part of their absorption spectra. Zeta-potential of Au NPs in this work is determined by the fraction of citrate anions on their surface. In the present work zeta-potential of laser-generated Au NPs is negative right after laser ablation, and no anions are needed. Moreover, Au nano-chains in [29] are not covered by a layer of $Ca(OH)_2$ and are spontaneously fragmented 20 days after synthesis.

Similar effect is also observed with another divalent cation, such as $Mg^{2+}$. Namely, laser ablation of a bulk gold target in aqueous solution of $MgSO_4$ leads to generation of elongated Au NPs. Details of these experiments will be presented elsewhere.

## Conclusions

It was experimentally shown that laser ablation of a gold target in water may lead to the formation of elongated gold nanoparticles. They are characterized by absorption in the red spectral region due to the longitudinal plasmon resonance, in addition to the standard peak of transverse plasmon resonance of gold in water in the vicinity of 520 nm. When a colloidal solution of elongated nanoparticles is exposed to pulsed laser radiation with intensity on the order of $10^9$ W $cm^{-2}$, agglomeration of elongated nanoparticles occurs in the initial stage. With a further increase in the exposure time, the aggregation of particles gives place to their fragmentation, i.e., an increase in the fraction of small spherical nanoparticles. The changes in the morphology of gold nanoparticles are confirmed by the extinction spectra of the colloidal solution and the TEM images. This change in the character of interaction of laser radiation with an ensemble of nanoparticles is typical specifically of elongated nanoparticles. Apparently, it was observed for the first time. The single-step technique of preparing elongated gold nanoparticles is of interest for biomedical applications because these nanoparticles absorb in the lasing band of most red sources used in photodynamic therapy.

## Acknowledgments

We are grateful to I.I. Rakov, K.O. Ayyyzhy, and S.V. Gudkov for their help in the experiment and Yu. L. Kalachev for his help in absorbance spectroscopy in Vis-IR region.

This work was performed within State Contract No. AAAA-A18-118021390190-1 and within the framework of National Research Nuclear University ‘MEPhI’ (Moscow Engineering Physics Institute) Academic Excellence Project (Contract No. 02.a03.21.0005) and supported in part by the Russian Foundation for Basic Research (Grant Nos 16-02-01054_a and 18-52-70012_e_Aziya_a, 18-32-01044_mol_a), Programme No. 7 of the Presidium of the Russian Academy of Sciences, and Grant MK 3606.2017.2 of the President of the Russian Federation for State Support of Young Russian Scientists.